**The Hidden Cost of Alloying: Disorder-Driven Transport Collapse in TMDs**

*Michele Pisarra, Clara Rebanal, Enrique Arévalo Rodríguez, Marc Meléndez, Elena Blundo, Giacomo Amadore, Jonathan J. Finley, Fabián Calleja, Marc G. Cuxart, Jesús Álvarez, María José Capitán, Fernando J. Urbanos, Julia García Pérez, Ramón Bernardo Gavito, Daniel Granados, Ji Dai, Massimo Tallarida, Antonello Sindona, Fernando Martín, Ferry Prins, Iolanda Di Bernardo*, Amadeo L. Vázquez de Parga**

M. Pisarra, A. Sindona
Dipartimento di Fisica Università della Calabria e INFN-gruppo collegato di Cosenza, Via P. Bucci, cubo 30 C, 87036, Rende, Italy

F. Calleja, Marc G. Cuxart, J. García Pérez, F. Jiménez Urbanos, R. Bernardo Gavito, D. Granados, F. Martín, A.L. Vázquez de Parga
Instituto Madrileño de Estudios Avanzados, IMDEA Nanociencia, Calle Faraday 9, 28049, Madrid, Spain.

C. Rebanal, E. Arévalo Rodríguez, M. Meléndez, F. Prins, J. Álvarez, I. Di Bernardo, A.L. Vázquez de Parga
Departamento de Física de la Materia Condensada, Universidad Autónoma de Madrid, Cantoblanco 28049, Madrid, Spain
Instituto de Física de la Materia Condensada IFIMAC, Univ. Autónoma de Madrid, 28049-Madrid, Spain
E-mail: iolanda.dibernardo@uam.es, al.vazquezdeparga@uam.es

E. Blundo, G. Amadore, J. J. Finley
Walter Schottky Institute, Technical University of Munich, 85748 Munich, Germany

M. G. Cuxart
Catalan Institute of Nanoscience and Nanotechnology (ICN2),CSIC and BIST, Campus UAB, Bellaterra 08193, Barcelona, Spain

J. Álvarez, I. Di Bernardo, A.L. Vázquez de Parga

Instituto de Ciencia de Materiales “Nicolás Cabrera”, Univ. Autónoma de Madrid, 28049-Madrid, Spain

J. Álvarez
Física de sistemas crecidos con baja dimensionalidad, UAM, Unidad Asociada al CSIC por el IEM, DP. 28006 Madrid, Spain

M.J. Capitán
Instituto de Estructura de la Materia IEM-CSIC, c/ Serrano 121, 28006 Madrid, Spain

J. Dai, M. Tallarida
ALBA Synchrotron, Carrer de la Llum 2-26 08290 Cerdanyola del Vallès, Barcelona, Spain

F. Martín
Departamento de Química, Universidad Autónoma de Madrid. 28049 Madrid. Spain

I. Di Bernardo
School of Physics and Astronomy, Monash University, Clayton, VIC 3800 Australia

**Abstract**

Alloying in two-dimensional semiconductors is widely used to tune bandgaps, yet its implications for charge and energy transport remain poorly understood. Here, we investigate $MoS_{2x}Se_{2(1-x)}$ alloys as a model system to study the interplay between composition, thickness, and disorder. Optical transitions and valence-band dispersions evolve continuously with both stoichiometry and number of layers, with negligible bandgap bowing and a composition-dependent attenuation of thickness-driven renormalization. In contrast, time-resolved spatial mapping of photoexcited carriers reveals a pronounced and asymmetric collapse of carrier diffusivity at intermediate compositions, which cannot be accounted for by changes in effective mass or band alignment, and instead emerges from strong real-space fluctuations in the local energetic landscape generated by random chalcogen substitution.

Microscopic simulations reproduce the experimental trends and show that the character of disorder depends critically on the direction of alloying, producing either scattering barriers or deep trapping sites. Together, these results demonstrate that transport in TMD alloys is governed by disorder physics, overlooked by conventional optical and photoemission probes at equilibrium. Our findings establish transport as a stringent metric of electronic quality and highlight intrinsic limitations in the usage of TMD alloys for layered semiconductor devices.

## 1. Introduction

Transition metal dichalcogenides (TMDs) have emerged as a versatile material platform for optoelectronics, excitonics, and layered semiconductor technologies, thanks to the combination of strong Coulomb interactions[1], spin–valley physics[2,3], and band structures that are highly sensitive to composition and dimensionality[4]. In particular, the possibility of tuning the electronic bandgap via both thickness reduction and alloying[5–7] enables virtually continuous tuning of optical transitions across a technologically relevant energy range. Among tuning strategies, isovalent alloying of TMDs promises relatively disorder-tolerant bandgap engineering without charge doping or the formation of structural phase boundaries, and has therefore attracted sustained interest. [8–12]

Chalcogen-mixed alloys such as $MoS_{2x}Se_{2(1-x)}$ have become a canonical model system for band structure engineering, as previous studies have shown that in the monolayer their optical bandgap evolves smoothly with composition[13]. Moreover, vibrational spectroscopies typically reveal well-defined dispersions and phononic modes across the full composition range, supporting the assumption that spectral tunability implies preserved electronic quality.[14] This virtually continuous spectral tunability—spanning modulation ranges of up to ~200 nm in wavelength[15]—underpins the widespread proposed use of transition metal dichalcogenide (TMD) alloys in device applications. For instance, carrier type conversion has been demonstrated in $WS_{2x}Se_{2(1-x)}$ nanosheet field-effect transistors (FETs)[16]. Enhanced electron transfer has also been observed in alloy-based heterostructures such as $MoSSe/MoS_2$ and $MoSe_2/MoSSe$, when compared to the pristine $MoSe_2/MoS_2$ system[17]. Furthermore, semiconductor-to-metal transitions have been achieved across a broad range of 2D TMD alloys, including $MoSe_{2x}Te_{2(1-x)}$, $WS_{2(1-x)}Te_{2x}$, $WSe_{2(1-x)}Te_{2x}$, and $Mo_{1-x}Re_xSe_2$ (see ref. [18] and references therein), among others.

Actually, energetic disorder strongly influences electronic transport, with theory predicting that high disorder can even drive a metal-insulator transition. In such systems, delocalized (conductive) states are separated in energy from localized (insulating) states by the mobility edge[19–21]. Experimental observations and material simulations confirm the emergence of such localized states upon introducing morphological or topological defects into the structure of 2D metals,[22] perovskites,[23] and carbon nanotubes.[24] Despite the widely accepted notion that crystal defects can degrade the performance of TMD alloy-based FETs[16,18], experimental insight into whether alloying intrinsically preserves or disrupts charge transport in otherwise high-quality TMD systems remains limited.

In this work, we analyse how composition and thickness affect the electronic structure and transport properties of $MoS_2$–$MoSe_2$ alloys. By a combination of microscopies, optical spectroscopy, angle-resolved photoemission and first-principles calculations we show that in our homogenous samples the bandgap evolves smoothly and adiabatically with both stoichiometry and thickness, with the thickness-induced changes being stronger toward the S-rich limit. Despite this apparent continuity in the single-particle electronic structure, carrier transport exhibits a profound and highly asymmetric degradation upon alloying. We demonstrate that random chalcogen substitution generates strong real-space energetic disorder that localizes or scatters carriers, even when the average band structure and excitonic response remain well defined. This decoupling between bandgap tunability and transport performance reveals a previously overlooked constraint in alloy-engineered 2D semiconductors, where energetic disorder becomes an invisible bottleneck for charge transport, thereby providing essential design rules for TMD-based optoelectronic and quantum devices that rely on both spectral control and efficient carrier propagation.

## 2. Results and discussion

***Composition and homogeneity.***

To establish the elemental ratio of the alloys, we performed a combination of synchrotron-based photoemission spectroscopy and Raman measurements on bulk samples with different nominal compositions. The results are reported in **Figure 1**.

We observe a systematic shift towards higher binding energy (BE) of the $S2p_{3/2}$ core level as a function of the Se content – from 160.14 eV for $MoS_2$ to 161.15 eV for the sample with the least S amount, visible in Fig. 1a. This shift, compatible with previous reports on similar TMD alloy[25,26] reflects the fact that S is smaller and more electronegative than Se, and as the S:Se ratio increases the electrons are more strongly pulled towards the S atoms, resulting in a shift towards lower BE. Notably, the photoionization cross-section of S2p is substantially higher than that of Se3p at our photon energy (330 eV), which accounts for the absence of discernible Se-related contributions in the spectral stack of Fig. 1a. The presence of Se in the sample is better elucidated in Fig. 1b, which presents the same dataset after background subtraction and normalization to the intensity of the $S2p_{3/2}$ peak, and a magnification on the intensity axis. Deconvolution of these spectra via peak fitting, repeated for different photon energies (330 and 500 eV) for consistency and yielding the same results, enables labelling according to stoichiometry: the $MoS_{2x}Se_{2(1-x)}$ samples analysed in the rest of this work have S:Se ratios of 2:0 ($MoS_2$), 1.5:0.5 ($MoS_{1.5}Se_{0.5}$), 1:1 (MoSSe), 0.85:1.15 ($MoS_{0.85}Se_{1.15}$), 0:2 ($MoSe_2$). Fig. 1c provides a quantitative representation of the $S2p_{3/2}$ BE position as a function of S concentration $x$. Energy-dispersive X-ray spectroscopy (EDX) composition estimation, collected in **Table S1**, yields similar results in terms of chalcogen ratios.

Complementary Raman spectroscopy analyses, shown in Fig. 1d, further substantiate these findings. The vibrational characteristics of the materials exhibit a strong dependence on the chalcogen ratio, in agreement with previous studies[26–29]. Specifically, the out-of-plane $A_{1g}$ and in-plane $E_{2g}^1$ phonon modes - known to follow single-mode and two-mode behaviors, respectively - are discernible for both parent compounds. The $MoS_2$ $A_{1g}$ mode undergoes a continuous redshift from 414 to 408 $cm^{-1}$ as $x$ decreases almost linearly from 2 to 0.85. Likewise, the $MoSe_2$-like $E_{2g}^1$ mode exhibits a systematic downshift with increasing S incorporation, indicative of bond softening effects driven by compositional modulation. The observed vibrational modes confirm that the materials are homogeneous alloy rather than Janus TMDs [30–32] or a clustering of different domains, as no additional peaks indicative of asymmetric chalcogen layering are observed.

To evaluate the homogeneity of the alloys, we performed a multi-scale structural and chemical characterization of MoSSe, virtually being the most disordered alloy and thus being selected as an exemplary sample. The results are reported in **Figure 2**.

The X-ray diffraction (XRD) data (Fig. 2a) indicate that MoSSe exhibits both short-range (atomic scale) and long-range ordering perpendicular to the basal plane, consistent with prior studies on 2D layered materials[33]. At high diffraction angles (short-range order) the pattern shows a limited number of peaks, with one dominant peak significantly more intense than the others, suggesting strong atomic ordering and a high degree of structural orientation. At lower diffraction angles (inset in Fig. 2a), oscillations in intensity are observed, indicating periodicity in the out-of-plane direction for the long-range order. The diffraction data correspond to a periodicity of approximately 25 nm, reflecting the presence of stacking faults in the bulk material. Scanning tunnelling microscopy (STM) measurements confirm structural homogeneity and high-level ordering, as seen in Fig. 2b: at atomic resolution, the lattice shows perfect periodicity and no defects or grain boundaries, and it is impossible to distinguish between the two chalcogens. While most physical parameters in TMD alloys (like the band edge position, see below) follow a quadratic law as a function of concentration, the lattice constant can be obtained quite accurately by linear extrapolation from the parent TMDs, as shown in our theoretical calculations in Figure S2 (Vegard's law[34]). The atomic lattice measured via STM (3.14 Å, see Fig. S2) is comparable to that expected for $MoS_2$ and $MoSe_2$.[28]
Scanning electron microscopy (SEM, shown in **Figure S3**) imaging of the MoSSe system reveals a morphology characteristic of a 2D-layered material, independent of its specific stoichiometry. The samples consist of large flakes exceeding 10 µm in lateral size, randomly distributed across the substrate. At higher magnifications, these flakes exhibit relatively smooth and continuous surfaces, a typical feature of exfoliated 2D materials. The absence of significant surface roughness or phase separation suggests a well-ordered structure without substantial aggregation or clustering of different elements. EDX measurements, performed on the Se and S elements and presented in Fig 2c, d respectively, show a homogeneous distribution of the chalcogen elements at the micrometer scale, indicating a uniform chemical composition rather than the formation of segregated domains.

***Adiabatic evolution of band structure.***
Photoluminescence (PL) measurements were carried out on all samples on exfoliated flakes of varying thickness from 1 to 10 layers; the results are collected in **Figure 3**. The flakes were exfoliated with the scotch tape method on $SiO_2$/Si substrates with $SiO_2$ thickness of 280 nm, and the flake thickness was identified based on optical contrast[35,36]. Fig. 3a reports representative PL spectra for monolayer and bulk-like (5 layers) flakes for all compositions examined. In agreement with previous experimental and theoretical works[37,38], the PL spectrum is dominated by the direct A exciton in the monolayer limit, while a transition to an indirect (I) exciton is observed for thickness larger than 2 layers in all the alloys (full dataset in **Fig. S4**). As shown in Fig. 3a, the A exciton (solid lines) in the monolayers exhibits a pronounced composition dependence, shifting towards lower binding energies by approximately 270 meV as a function of $x$, whereas the indirect I exciton (dashed lines) shows a much smaller shift of only about 40 meV. In contrast, the I exciton shows a strong thickness dependence, as reported in panel 3b for the exemplary $MoS_{0.85}Se_{1.15}$, shifting downwards by about 200 meV from 2 to 5 layers before remaining nearly pinned at higher layer numbers. By comparison, the A exciton displays only a negligible shift with thickness. The large (small) composition-dependence and small (large) thickness-dependence of the A (I) exciton can be explained by the nature of the electronic wavefunctions in the system: The states at the **K** point of the Brillouin zone (BZ), in both the conduction and valence bands, are involved in the A exciton and consist mainly of the metal *d* orbitals and the chalcogen $p_x$, $p_y$ orbitals (these, lying lower than the metal ones). Due to their in-plane character, these states are atomically localized within each layer and do not extend into the interlayer space. In contrast, the states at the **Γ** point (valence band) and **Λ** point (conduction band), which contribute to the I exciton, are predominantly composed of the metal *d* orbitals and the chalcogen $p_z$ orbitals. Their out-of-plane orientation allows them to extend significantly into the interlayer region.[39] While bandgap tuning was previously reported for thick films[29,40] or mono[13]- and bi-layer samples,[27] our extensive dataset allows us to establish generalized rules on the grounds of orbital composition: while composition sets the dominant energy scale for the A exciton, interlayer coupling contributes with a composition-dependent correction; the opposite is true for the I exciton.

Ternary TMD alloys are expected to exhibit bandgap bowing as a function of both composition and number of layers[41–43], with the bandgap obeying the equation:

$$E_g(x,N) = xE_g(0,N) + (1-x)E_g(1,N) - \beta x(1-x) \qquad (1)$$

where $E_g$ is the bandgap size, *x* the concentration of one of the chalcogens, *N* is the number of layers and $\beta$ is the bowing parameter. $\beta$ depends on the size and electronegativity differences between the two constituent compounds, being largest for compounds with mixed transition metals (like $Mo_xW_{(1-x)}S_2$) and smallest for S-Se alloys [34,43]. As demonstrated in Fig. 3c (see also **Fig. S5** and **Table S2** for full data sets), we observe the bowing vs. composition to be negligible across different thicknesses within experimental resolution, both in the monolayer where the gap is direct (Fig 3c, diamonds) and in the thicker samples (Fig. 3c, circles). Overall, a smooth and quasi-linear dependence of the A exciton energy on composition is observed.

To support our experimental findings, we calculated (see Experimental Section/Methods section for details) the optical absorption spectra for $MoS_2$, $MoS_{1.5}Se_{0.5}$, MoSSe, $MoS_{0.5}Se_{1.5}$, $MoSe_2$ as a function of varying thickness. The results are collected in **Fig. S6.** For monolayers, where confinement is known to result in the shifting of the bands at the **K** point above the valence band maximum (VBM) at Γ (indirect to direct bandgap transition[44,45]) this also corresponds to the fundamental gap. From these calculations we extract an estimation of the minimum optical gap, which in turn gives information on the size of the direct gap – i.e., the minimum vertical distance between the valence and conduction bands across the BZ. These values are obtained by estimating the onset of the first absorption plateau and are reported in Figure 3d. We note that the calculations predict a modest thickness-dependent reduction of the direct gap, whereas the experimental PL data show only a weak variation beyond the monolayer; this discrepancy likely arises from the different nature of the probed quantities, as the A-exciton energy includes excitonic and screening effects that are not fully captured in the present single-particle optical absorption analysis. Additionally, the exciton binding energy is expected to vary with composition, being larger for $MoSe_2$ than for $MoS_2$,[46] compensating the already small bowing predicted at the single particle level.

To evaluate which electronic states are responsible for the observed thickness and composition changes to the electronic structure, we turn to the valence band mapping of $MoS_{2x}Se_{2(1-x)}$ alloys as a function of chemical composition, reported in **Figure 4**: panels 4a-e illustrate the band dispersion along the $\bar{\mathbf{M}} - \bar{\mathbf{\Gamma}} - \bar{\mathbf{K}}$ high-symmetry direction, obtained with LH polarization for all samples. Experiment geometry is reported in **Figure S7**.
Spectral features appear sharper for the parent TMD samples and significantly more blurred for the alloys: this is a consequence of the compositional disorder of the latter, resulting in a larger Gaussian contribution to the lineshapes and overall broader bands. A key observation is that across chemical compositions the valence band maximum (VBM) at $\bar{\mathbf{\Gamma}}$ remains stationary, while the VBM at $\bar{\mathbf{K}}$ shifts systematically towards higher binding energies with increasing $x$, changing the distance between the valence band maxima at $\bar{\mathbf{\Gamma}}$ and $\bar{\mathbf{K}}$ (Fig. 4f, blue line and right-handside scale). Additionally, we observe the spectral weight at $\bar{\mathbf{\Gamma}}$ on the top band to decrease with increasing Se content, whereas deeper-lying bands gain intensity. Concomitantly, the energy splitting of the valence bands at the $\bar{\mathrm{K}}$ point (L1 and L2 bands as marked in Fig. 4a) decreases systematically with $x$ (Fig. 4f, left-hand axis). These trends indicate a composition-dependent enhancement of the $\bar{\mathrm{K}}$ -point spin splitting, consistent with a modulation of the out-of-plane spin–orbit field arising from the atomic spin–orbit coupling responsible for the splitting[47].
The observation of well-defined ARPES spectra for the alloys and the experimental validation of the linearity of the trends as a function of composition confirm that the electronic shells change shape adiabatically, validating the wide-held assumption that the position of the band edges of these samples can be fine-tuned by controlling stoichiometric ratios.
Previous theoretical works[43] showed that the highest occupied molecular orbital (HOMO) and the lowest unoccupied molecular orbital (LUMO) states in TMD-ML are to be mainly attributed to the metal atoms and in lesser part to the chalcogens, and follow the symmetry of the irreducible representations A1' (singlet $d_{z^2}$ state) and E' (doublet of $d_{x^2-y^2}$ and $d_{xy}$ states) of the point group D3h, respectively. To support and understand our experimental findings, we calculate the atom-projected DOS for the studied alloy systems (Fig. 4g-h). Regardless of the chalcogen composition, the highest pDOS weight is on the Mo atoms in both the valence band top and in the conduction band bottom, justifying the smooth variation of spectral features.

***Carrier transport.***

Our experimental and theoretical results so far consistently indicate that $MoS_{2x}Se_{2(1-x)}$ alloys behave as electronically well-defined, adiabatically tunable semiconductors - vibrational modes evolve smoothly with composition, and both optical and photoemission measurements reveal a continuous modulation of band edges and orbital character with no signatures of phase segregation or abrupt electronic transitions. From the perspective of single-particle electronic structure, alloying therefore appears as a benign and controllable tuning parameter.

Carrier transport results (**Figure 5**), however, reveal a striking departure from the otherwise smooth evolution observed in spectroscopic probes, with results that are highly composition-dependent and cannot be explained by changes in effective mass or band-edge alignment alone. We performed transient scattering microscopy on mechanically exfoliated flakes of both pure and mixed phases. The operating mechanism for this technique is explained in previous reports.[48–51] In short, by using an above-bandgap pump and a widefield probe, this method tracks local refractive index changes induced by photoexcited states, allowing us to map their spatial evolution with sub-nanosecond precision. This yields a time-dependent expansion of the carrier distribution, indicating fast carrier diffusion. We then quantified carrier transport by computing the spatial variance of azimuthally averaged transient scattering images following a discrete-method calculation.[52]. We obtain the values of the initial diffusivity by performing a linear fit to the first 1.5 ns, before conventional trapping mechanisms lead to a deceleration of carrier transport. Exemplary data for $MoSe_2$ are collected in **Figures 5a-c**.

In contrast to pure $MoSe_2$ and $MoS_2$, non-pristine $MoS_xS_{(2-x)}$ phases revealed a rapid, composition dependent suppression of carrier diffusion during the initial diffusive regime. Notably, this dramatic reduction in diffusivity occurs in the absence of any discontinuity in the optical bandgap or valence-band dispersion measured by PL and ARPES, underscoring a fundamental decoupling between spectroscopically inferred band structure and transport behavior. Indeed, numerical simulations predict that substituting random chalcogen sites makes the crystal's energy landscape highly inhomogeneous, even at low levels of mixing (10% and 90% sulfur content) as seen in Fig 5d. Critically, doping of sulfur atoms in $MoSe_2$ leads to a completely different potential landscape compared to the inverse. For compositions closer to $MoSe_2$, the phase with lower bandgap, replacing Se atoms with S leads to the formation of high energy sites, which act as effective scattering centers for the impinging carriers, increasing the total number of scattering events and thus, reducing the total mobility. For compositions close to $MoS_2$, on the other hand, the introduction of Se atoms creates low energy sites that act as deep trapping sites, completely stopping carrier diffusion altogether. This implies that for compositions with high $x$ the diffusivities will be suppressed much rapidly compared to low values of $x$. Indeed, in Fig. 5e, both experimental observations (black dots) and simulations (blue line/shaded area) align with this hypothesis, strongly suggesting that the origin of the suppressed transport arises from energetic disorder.

The key to reconciling the transport results with the ones reported before lies in the nature of the disorder introduced by alloying. While the delocalized Bloch states forming the band edges evolve smoothly with composition, random substitution of chalcogen atoms creates strong real-space fluctuations in the local potential[53]. These fluctuations generate energetic barriers and traps that efficiently scatter or localize carriers, even when the average band structure remains well-defined. In this sense, transport provides a complementary — and more stringent — metric of electronic quality than bandgap tunability alone, as it is the first measurement sensitive enough to collapse under disorder.

Interestingly, similar observations have previously been reported for mixed halide perovskites,[54] where halide mixing is accompanied by a clear spectral broadening in PL spectroscopy measurements,[55] due to the of energetic disorder created by the higher/lower energy sites where the halides are introduced.

## 3. Conclusion

In this study, we present a framework to understand how composition and thickness govern the electronic properties of TMD alloys. We demonstrate that the optical bandgap and the valence-band structure evolve smoothly and adiabatically across the full composition and thickness range. Importantly, we disentangle the effects of thickness and composition on the A and I excitons observed in PL experiments and explain them in light of orbital geometry. Time-resolved carrier-transport measurements reveal a dramatic and asymmetric suppression of diffusivity in mixed compositions, despite the absence of abrupt changes in the average band structure. Supported by numerical simulations, we attribute this behavior to strong real-space potential fluctuations arising from random chalcogen substitution, which generate scattering barriers or deep trapping sites depending on the direction of alloying. Together, our results show that while chemical composition and thickness provide powerful and continuous control over the electronic structure of TMD alloys, carrier transport is governed by a fundamentally different principle: the emergence of disorder-induced energetic landscapes that are invisible to conventional spectroscopic probes. More broadly, our findings identify an intrinsic limitation of alloy-based bandgap engineering in two-dimensional materials and establish design rules for optoelectronic and quantum devices that require both spectral tunability and efficient carrier propagation.

**4. Experimental Section/Methods**

*Sample preparation*: Samples with different Se:S stoichiometric ratios were purchased from HQ graphene, and exfoliated in ultra high vacuum (UHV) environment prior to UHV-based measurements (ARPES, SEM, STM). For optical measurements, thin flakes were mechanically exfoliated by the scotch tape method on SiO2/Si substrates with $SiO_2$ thickness of 280 nm. Hundreds of thin flakes were exfoliated in total from all the alloys and parent materials. A color contrast analysis was employed to identify the flake thickness.

*STM:* STM measurements were carried out in a low-temperature STM operating in Ultra-High Vacuum at a base pressure in the low $10^{-10}$ mbar. The UHV system is equipped with a preparation chamber and a load lock that allows sample exfoliation and transfer to the STM without breaking the UHV conditions. During measurements the sample temperature was kept at 77K, as lower temperatures would hinder enough carrier mobility in the samples to allow for tunnelling.

*XRD*: Powder X-ray diffraction data were collected using a Siemens D5000 diffractometer equipped with a Cu anode for the structure determination of the samples. The X-ray source is equipped with a horizontal soller slits setup to reduce the horizontal divergence. The diffractometer was configured in the θ-2θ configuration. The diffracted signal was recorded using a NaI(Tl) scintillator from Oxford Instruments combined with an automatic attenuator device[56] to increase its dynamic detection range. The detector arm was equipped with a soller slit and a graphite monochromator to suppress the signal due to fluorescence and the CuKβ line. To select useful events, a tunable differential discriminator with a reconfigurable energy window was used with the amplified detector signal. The energy window was selected with the aid of a multichannel analyzer. The sample was placed on a goniometer head equipped with two circles for normal to the surface alignment. The surface alignment was made with the aid of a laser reflected beam and later on maximized with the perpendicular to the surface [002] bulk Bragg peak. A motorized precision z-axis movement was used for the correct placement of the sample in the beam. The collected diffraction data corresponded to both Cu Kα1,2 lines; the contribution of the Cu Kα2 line was later removed from the recorded data by a mathematical procedure using a fit with penalized splines[57]. The obtained diffraction pattern was indexed using the N-TREOR9 software[58] integrated in the EXPO2014 package.[59] The atom position was determined by direct methods followed by a final step Rietveld refinement.

*Raman measurements:* Raman measurements were taken using an optical spectroscopy set-up comprising a 0.5 m Andor spectrometer with a back-illuminated Si CCD camera. The excitation source used was the 488 nm line of an Ar ion CW laser. The samples were introduced in an Attodry 800 optical cryostat that allows for temperature control between 4K and room temperature. The cryostat is equipped with nanometric precision positioners and scanners to navigate the samples and perform spectral mapping.

*SEM/EDX measurements:* Scanning electron microscope (SEM) images and Energy-dispersive X-Ray Spectroscopy (EDX) were carried out into a high-definition SEM (Sigma acquired from Carl-Zeiss) equipped with a Peltier cooled EDS detector (Xflash 430 from Bruker), with 30 $mm^2$ detector area, 133 eV resolution and a working temperature of -20 °C. EDX element mappings were acquired with an acceleration voltage of 10 kV and an acquisition time of 10 min.

*PL measurements:* Micro-PL measurements were acquired by exciting samples with a 520-nm diode laser (by Cobolt). The PL signal was spectrally dispersed by a 0.5 m focal length spectrometer (HRS500 by Princeton Teledyne) equipped with a 300 grooves/mm and was detected by a back-illuminated etalon-free Si CCD camera with enhanced quantum efficiency at 1000 nm (Blaze HRX100 by Princeton Teledyne). The laser light was filtered out by a long-pass filter at 550 nm (by Thorlabs). A 100× objective with NA = 0.81 (by Attocube) was employed to excite and collect the light, in a backscattering configuration using a confocal setup.

*ARPES measurements*: The measurements in this work were carried out at the LOREA beamline at ALBA Synchrotron using an MBS A1 hemispherical energy analyser. The samples were kept at 90K during acquisition to avoid charging.

*Transport measurements*: Transient scattering microscopy was performed in a home-built microscope using the same configuration as [52]. In short, a combination of two pulsed picosecond laser diodes is used for this pump-probe based experiment, with an above bandgap 405 nm diode as the pump (Picoquant LDH-DC-405) and a 780 nm as a probe (Picoquant LDH-DC-780). Both lasers are controlled via the same driver (Picoquant Sepia II) which allows control of the delay between the two lasers via its electronic capabilities. After spatially filtering both lasers the two are combined using a dichroic mirror and sent to the objective (Nikon Plan Apo 1.45 NA 100x). The reflected light is sent back to a CMOS camera after filtering out remaining pump light. The pump laser is modulated at 440 Hz, while the CMOS acquires pictures at twice the speed (880Hz), thus acquiring consecutive images with and without the pump excitation. We analyze our images by dividing subsequent ON/OFF image pairs, thus obtaining differential images *Diff = ON/OFF-1*, which contain information about how much the refractive index of the material changes after a photoexcitation. A single experiment typically involves 4000 pairs for each pump-probe delay, the pairs are averaged together to reduce background noise as much as possible.

*Numerical simulations: :* To determine how excitons travel through different energetic land-scapes we simulate the process as a series of Brownian walkers interacting with the ions present in the crystal lattice. Each exciton follows a random path described by a stochastic differential equation, in the Itô interpretation.

$$\Delta r = \frac{D}{k_B T} F \Delta T + \sqrt{2D_0} dW \quad (2)$$

Where $D_0$ is the diffusion coefficient, $dW$ is taken from a Wiener process such that $< dWdW > \Delta t$ and $F$ is the force felt by an exciton given by the energy potential ($F = -\nabla V$). The potential $V$ can be calculated such as

$$V(r_i) = \Delta E \frac{\sum_j I_j u(r_{ij})}{\sum_j u(r_{ij})} \quad (3)$$

Where $\Delta E$ is the bandgap difference, $I_j$ equals 1 if there is a replaced atom at $r_i$ and 0 otherwise and $u_{ij}$ is a Gaussian weighted function to simulate the effect of the landscape in the excitons ($u(r) = \sqrt{2/_{\pi a}}\, e^{-2r/_a}$ , where the distance $a$ represents the distance over which an ion has influence). Substituting,

$$F(r_i) = -\frac{2\Delta E}{a} \frac{\sum_j I_j u(r_{ij}) \widehat{r_{ij}}}{\sum_j u(r_{ij})} + \frac{2\Delta E}{a} \frac{\sum_j I_j u(r_{ij}) \sum_j I_j u(r_{ij}) \widehat{r_{ij}}}{(\sum_j u(r_{ij}))^2} \quad (4)$$

We describe the crystal as a grid of 250x250 unit cells with periodical boundary conditions. And let the excitons diffuse following eq. (2).

To find the energetic landscape of each composition we randomly fill chalcogen atom sites with either S or Se. By finding the convolution of the S sites with the carrier particle probability function we calculate the local concentration of S sites that affect the carrier, thus finding the local bandgap for each composition. Brownian motion simulations are performed by introducing a Gaussian-shaped diffraction limited excitation at the center of these grids and tracking the motion of carriers as time passes.

*DFT Methods:* Density Functional Theory (DFT) calculations were carried out within the Projector Augmented Wave (PAW) method[60] as implemented in the VASP [61–63] In all calculations we employed the PBE exchange-correlation functional[64] and included the Grimme's D3 dispersion corrections[64], which has proven to be effective in reproducing both the in-plane and the out of plane lattice constants of $MoS_2$ and $MoSe_2$[45]. The plane-waves cutoff energy was set to 400 eV; the self consistent cycle tolerance to $10^{-6}$eV; the 1st Brillouin Zone integrations were carried out adopting $\Gamma$ centered Monkhorst-Pack grids[65] of $(12 \times 12 \times 5)$ and $(12 \times 12 \times 1)$ for the bulk and slab materials, respectively (more refined grids were used for the computation of the density of states and the light absorption). A $5 \cdot 10^{-3}$ eV/Å convergence criterion on the maximum residual force was adopted in all geometry optimization. Finally, a 20 Å vacuum region between system replicas was used in the slab calculations.

In the hexagonal phase the $MoX_2$ material (X=S or Se) has a layered structure. In each single layer a plane of Mo atoms is sandwiched between layers of X atoms and each molybdenum atom makes coordination bonds with the 6 chalcogen atoms in a trigonal prismatic geometry. The 2H bulk phase is obtained stacking $MoX_2$ single layer in the so-called AA' stacking[65], where the Mo atom of the top layer is placed directly above the X atom of the bottom layer, and vice versa. Hence, the 2H-$MoX_2$ bulk material has a minimal unit cell containing 2 Mo atoms and 4 chalcogen atoms. In our analysis, we constructed an in-plane $(\sqrt{3} \times \sqrt{3})R30°$ supercell, which is characterized by a total of 6 Mo atoms and 12 chalcogen atoms, and a total of $2^{12} = 4096$ possible mixed composition Mo-S-Se arrangements. The latter are reduced to 179 inequivalent arrangements (see the SI for more details) characterized by a S:Se ratio of 2:0 ($MoS_2$), 11:1, 5:1, 3:1, 2:1, 7:5, 1:1, 5:7, 1:2, 1:3, 1,5, 1:11, 0:2 ($MoSe_2$). Given a specific initial atomic configuration, a geometry optimization was carried out relaxing the coordinates of all the atoms and optimizing the lattice vectors. Whenever multiple arrangements belonged to the same composition, the one realizing the minimum energy was chosen for further analyses and for the creation of the initial geometries of the slab calculations.

Light absorption is calculated as the imaginary part of the dielectric function $\epsilon_2(q,\omega)$ in the long wavelength limit ($q \rightarrow 0$) and neglecting local field effects. $\epsilon_2$ is obtained through a Linear Response TD-DFT calculation within the PAW based implementation of the VASP package[66]. For these calculations we set a very strict $10^{-8}$ eV energy tolerance and included enough bands to cover the energy region up to at least 20 eV above the Fermi level.

**Supporting Information**

Supporting Information is available from the Wiley Online Library or from the author.

**Figures**

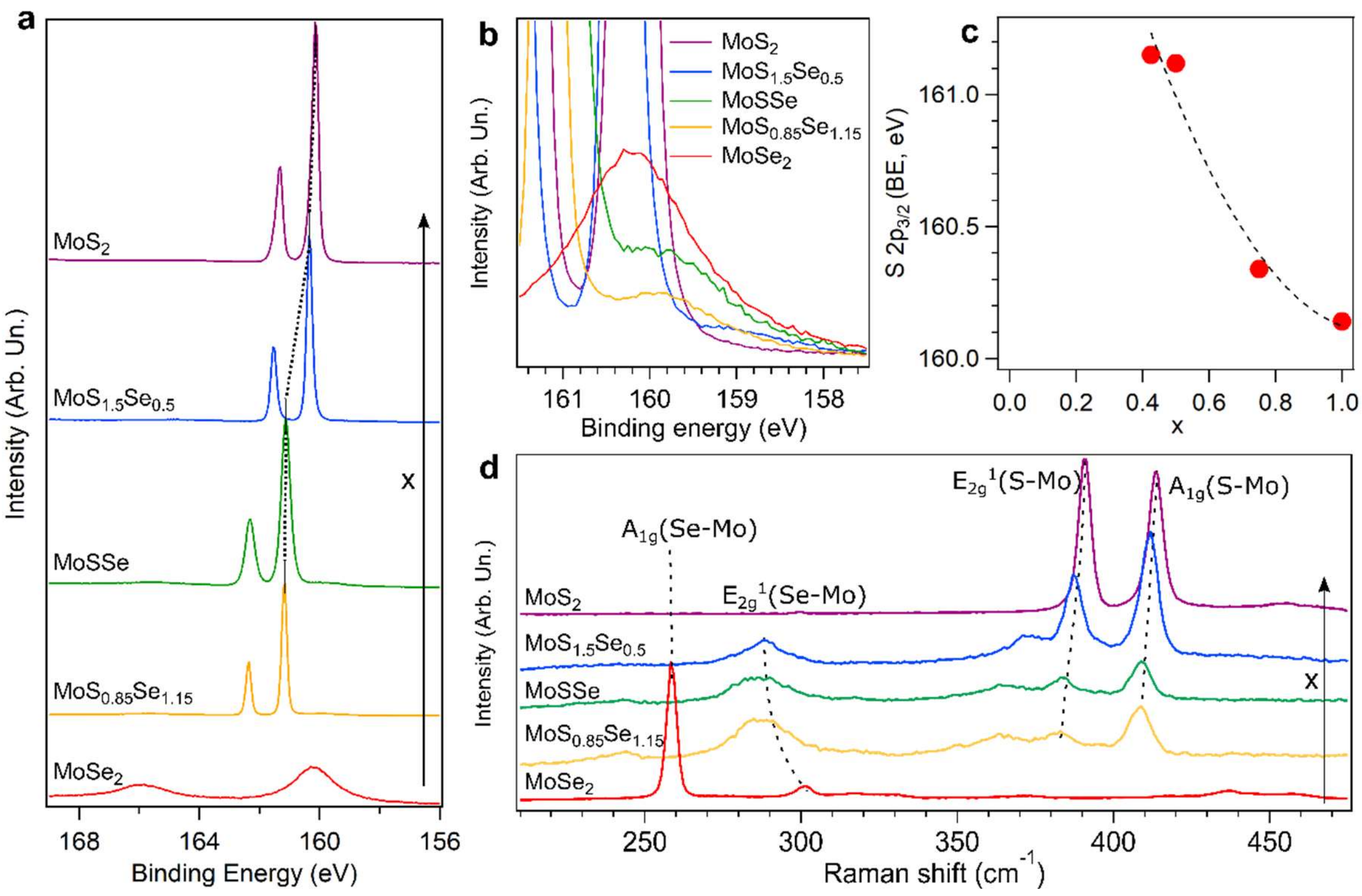

*Figure 1* Spectroscopic characterization of $MoS_xSe_{2(1-x)}$ samples. (a) S2p and Se3p core levels, photon energy 330 eV. Spectra are vertically offset for clarity; dashed lines are a guide for the eye to help tracking the position of the $S2p_{3/2}$ component. (b) Close-up on the Se3p core level, spectra are normalized to the $S2p_{3/2}$ intensity and background is subtracted. (c) Evolution of the $S2p_{3/2}$ component centroid as a function of sulphur content *x*. The dashed parabolic line serves as a guide for the eye. (d) Raman spectra of all samples, vertically stacked and offset for clarity. Dashed lines are guides for the eye.

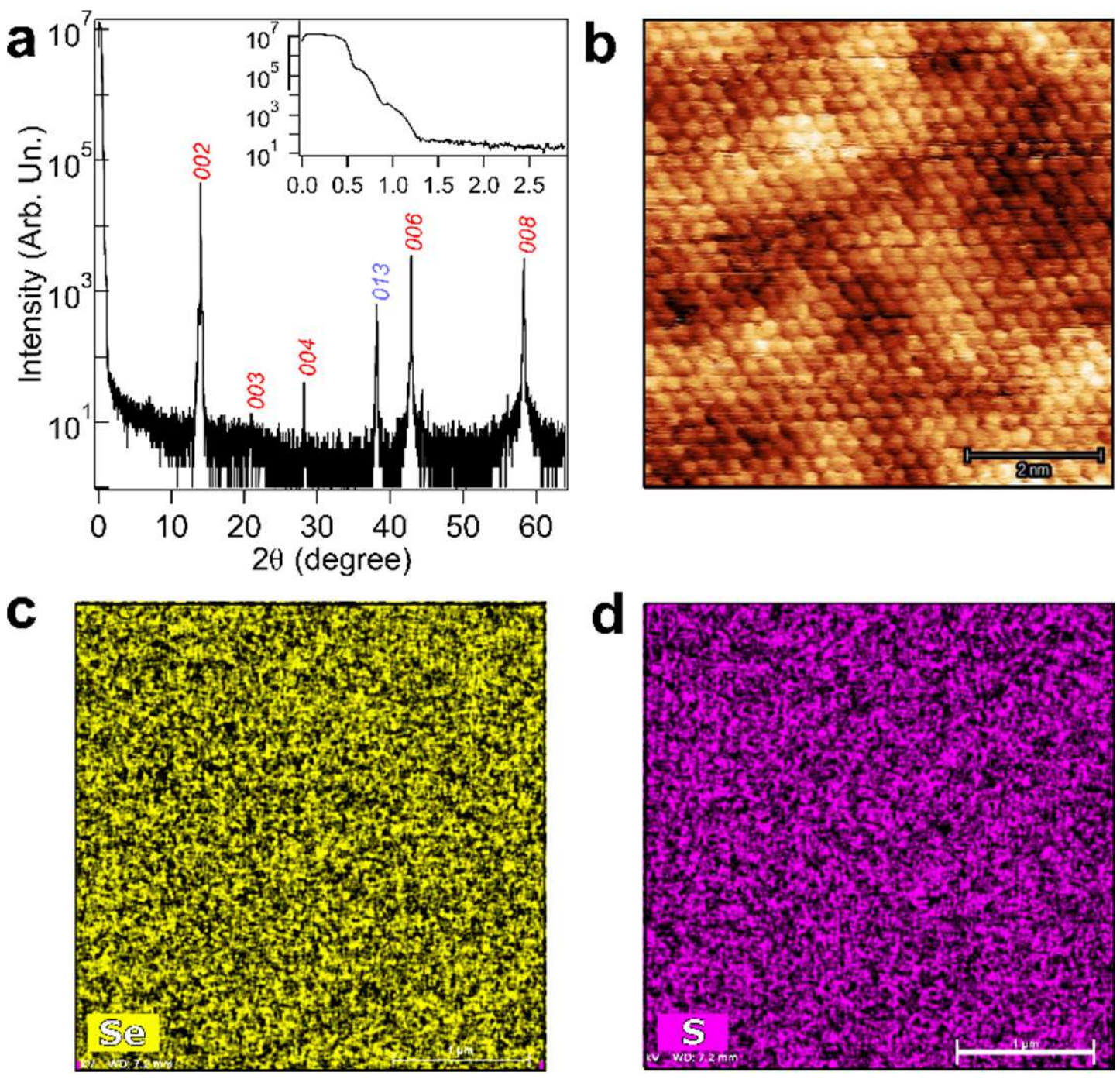

*Figure 2* Structural characterization of MoSSe. (a) XRD in the direction perpendicular to the plane; inset: close up at small diffraction angles. (b) Atomically resolved (10x10) nm STM image of MoSSe; I=200 pA; V=3,2V. (c), (d), EDX mappings of MoSSe on the Se-L transition and S-K transition respectively.

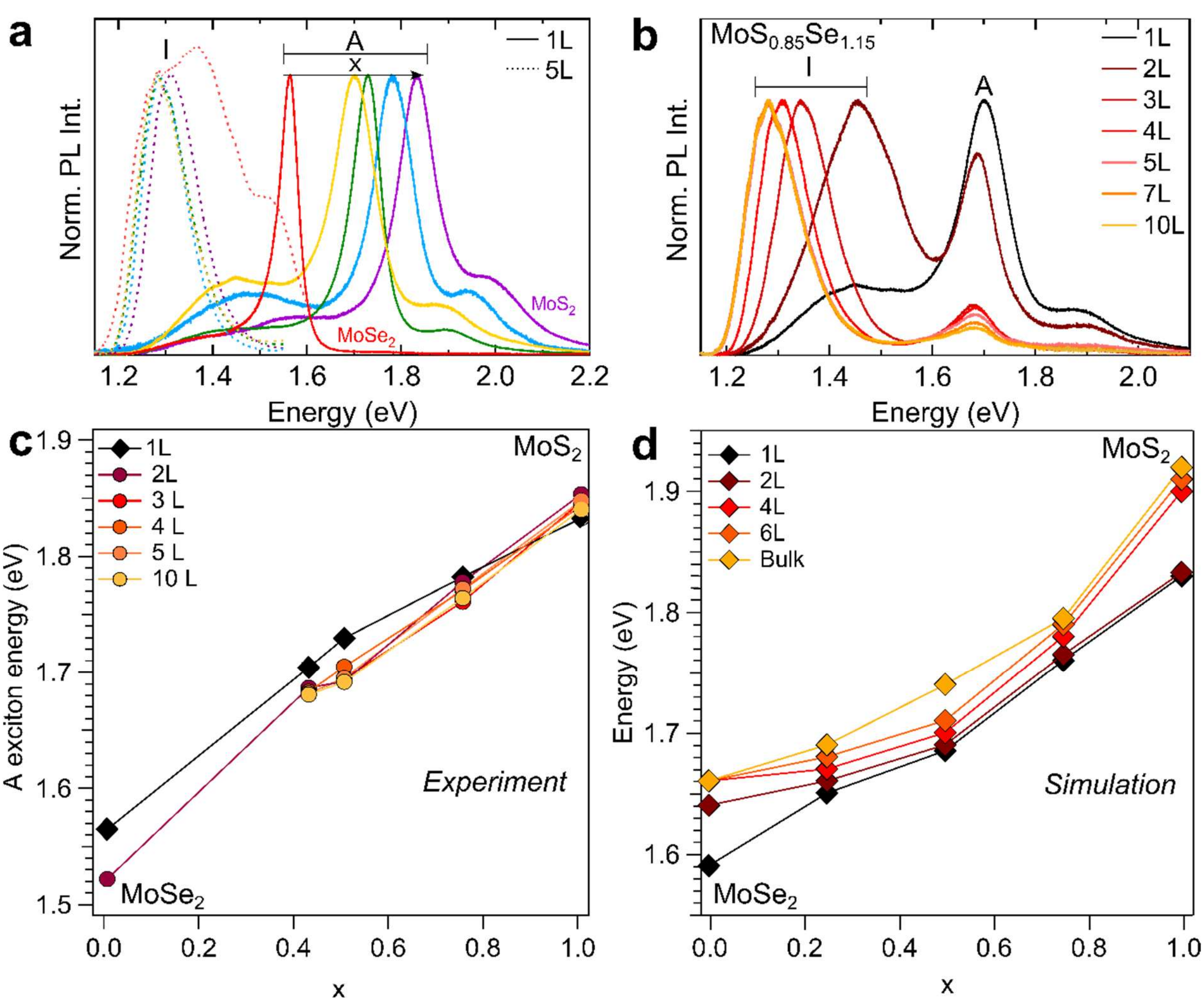

*Figure 3* (a) PL spectra for monolayers (1L, solid lines) and bulk-like flakes (5L, dashed lines) of all samples. Spectra have been normalized to the intensity of the A exciton (for 1L) or I exciton (for 5L). (b) Normalized PL spectra for a selected alloy (with S:Se = 0.85:1.15) as a function of thickness. The 1L spectrum is dominated by the A exciton, while for ≥2L the I exciton dominates. (c) Optical bandgap size (A exciton energy) obtained by PL spectra for different sample thicknesses and as a function of composition. Diamonds (circles) mark a direct (indirect) bandgap. (d) Theoretical variation of the electronic direct bandgap for samples with different S:Se ratio for different number of layers and as a function of S concentration.

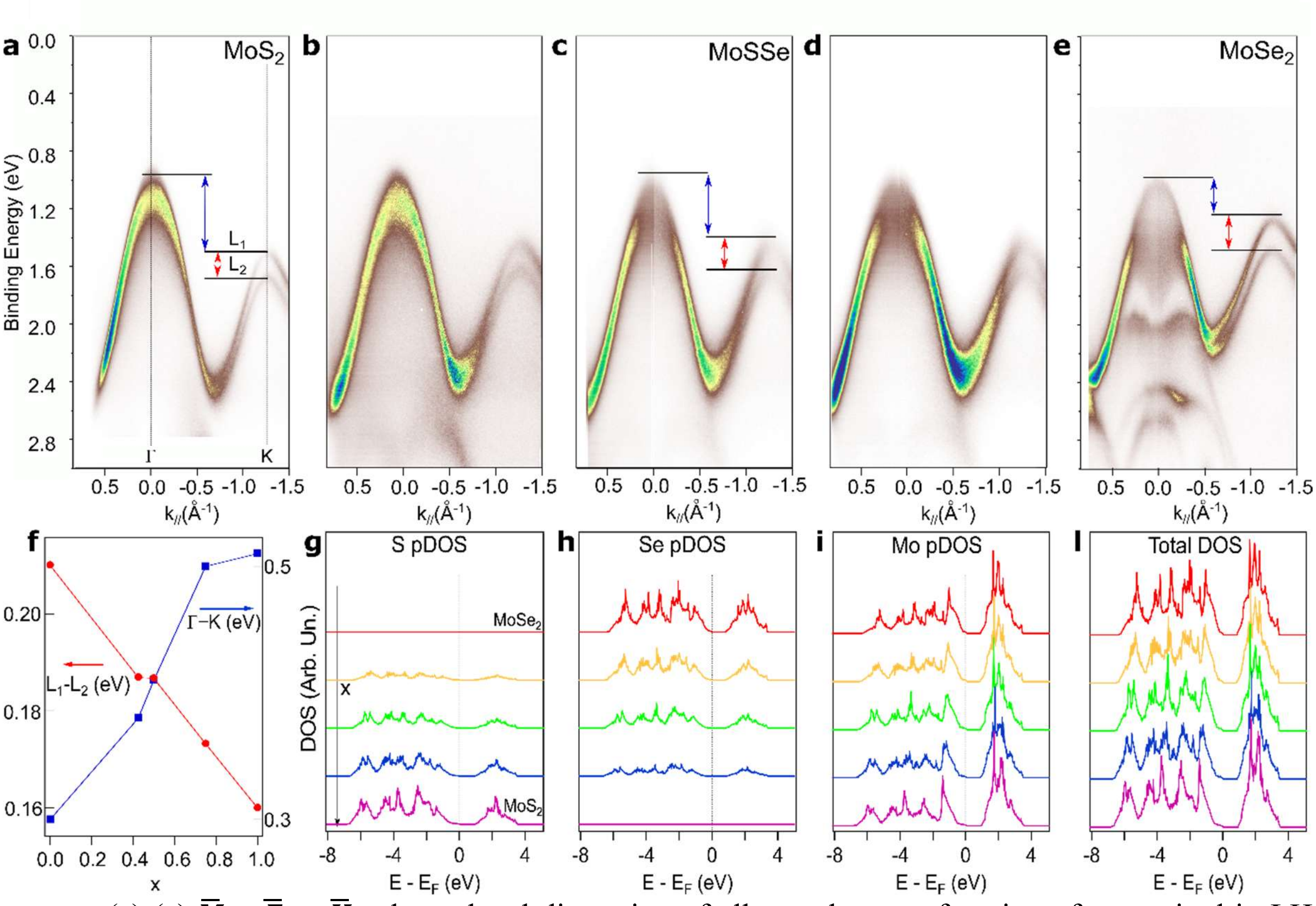

*Figure 4* (a)-(e) $\overline{\boldsymbol{M}}-\overline{\boldsymbol{\Gamma}}-\overline{\boldsymbol{K}}$ valence band dispersion of all samples as a function of $x$ acquired in LH polarization. (f) Band splitting at the $\overline{\boldsymbol{K}}$ point (red line and circles, LHS vertical scale) and distance between the band maxima at $\overline{\boldsymbol{\Gamma}}$ and $\overline{\boldsymbol{K}}$ point (blue line and squares, RHS scale) for all samples as a function of $x$. (g), (h), (i) pDOS on the S, Se, and Mo atoms and total DOS (l) as a function of composition; spectra are vertically offset for clarity.

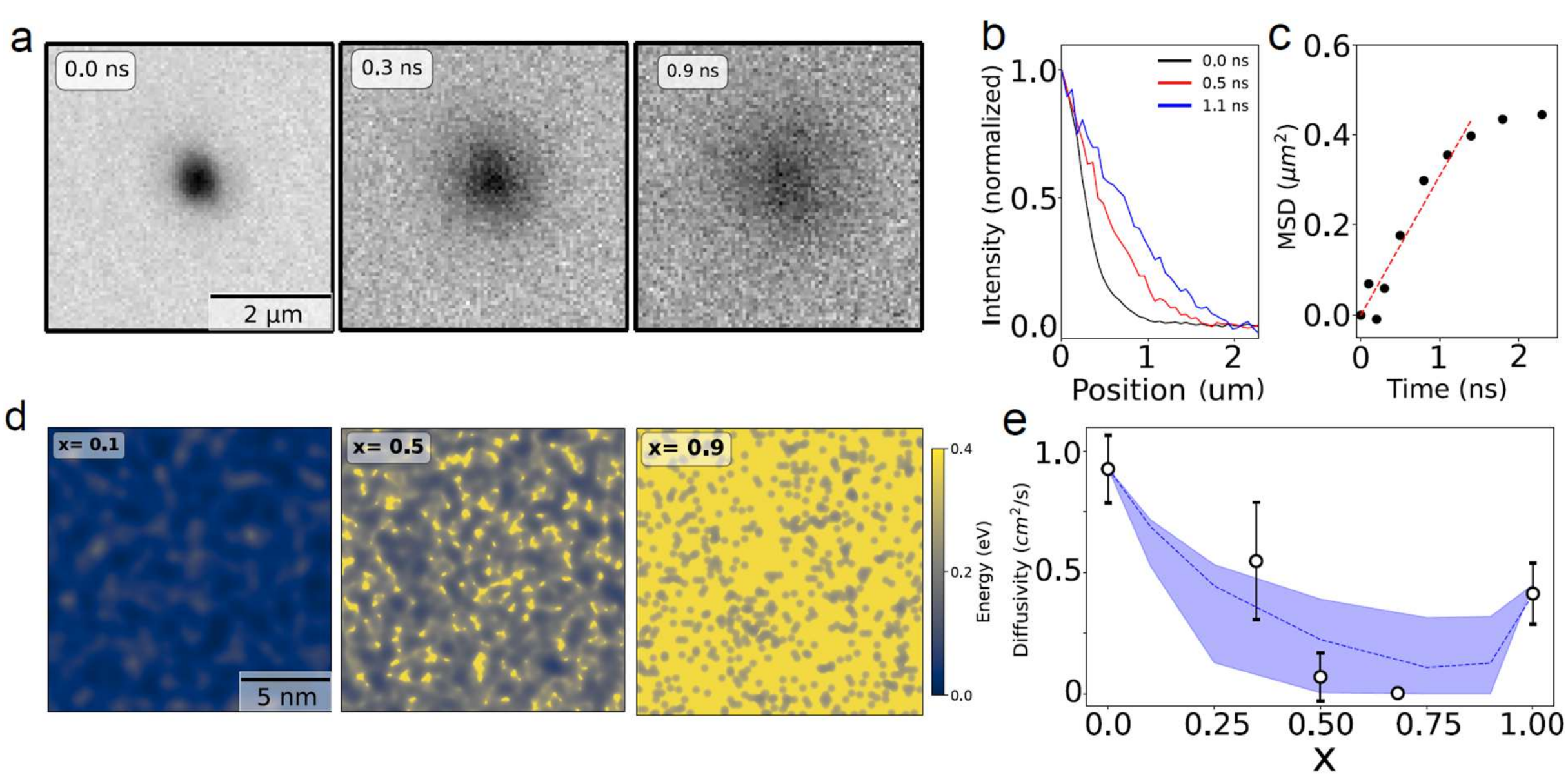

*Figure 5* (a) Exemplary transient scattering images collected for pure $MoSe_2$ at different pump-probe time delays. (b) Resulting azimuthal averaged profiles from the dataset shown in (a). (c) Extracted variance and linear fit to calculate the value of diffusivity, the fit is performed on the first 1.5 ns after photoexcitation. (d) Simulated energetic landscapes for $x = 0.1$, 0.5 and 0.9 (e) Extracted values of the diffusivity for experiments (black points) and simulations (shaded blue area). The simulation area corresponds to different parameters in the ion Bohr radius, from $a_{Se} = 8.2$ to 1 Å and $a_S = 1$ to 0.9Å.

**Acknowledgements**

The authors kindly acknowledge Prof. A. Polimeni and M.A. Valbuena for insightful discussions. I.D.B. acknowledges funding from the European Union's Horizon Europe research and innovation programme under the Marie Skłodowska-Curie grant agreement no.101063547 (STORM) and support from the Ramón y Cajal program, grant no. YC2022-035562-I. ALVP acknowledges funding from the Ministerio de Ciencia, Innovación y Universidades (MICIU/AEI/10.13039/501100011033) through grant PID2021-128011NB-I00. ALVP and F.P. acknowledge Comunidad de Madrid through grant "Materiales Disruptivos Bidimensionales (2D)" MAD2D-CM-UAM funded by the Recovery, Transformation and Resilience Plan, and by NextGenerationEU from the European Union. IMDEA Nanociencia and IFIMAC acknowledge financial support from the Spanish Ministry of Science and Innovation 'Severo Ochoa' (Grant CEX2020001039-S) and 'María de Maeztu' (Grant CEX2018000805-M) Programme for Centers of Excellence in R&D, respectively. LOREA was co-funded by the European Regional Development Fund (ERDF) within the Framework of the Smart Growth Operative Programme 2014-2020. M. P. and A. S. acknowledge partial financial support by the Centro Nazionale di Ricerca in High-Performance Computing, Big Data and Quantum Computing,

PNRR 4 2 1.4, CI CN00000013, CUP H23C22000360005. We acknowledge allocation of computing time at the Centro de Computación Científica of the Universidad Autónoma de Madrid and at the 'Leonardo' and 'Galileo100', facilities of the CINECA Consortium, within the INF16-npqcd project under the CINECA-INFN agreement and the 'Newton' and 'Alarico' high performance computing clusters, provided by the University of Calabria. E.B. and J.J.F. gratefully acknowledge the German Science Foundation (DFG) for financial support via the Cluster of Excellence Munich Center for Quantum Science and Technology (MCQST, EXC2111). E.B. gratefully acknowledges support from the Alexander von Humboldt Foundation through a Humboldt Research Fellowship. F.P. acknowledges funding by the European Union (ERC, En-Vision, project number 101125962), from the Spanish AEI (PID2022-141579OBI00). F.M. acknowledges financial support from the Spanish Ministry of Science and Innovation, Grant no. PID2022-138288NB-C31 (ESTUDYAMOS). D.G. and R.B.G. acknowledge the financial support from the Spanish Ministry of Science and Innovation through the project PID2022-137779OB-C42 (MAINSTREAM). J.G.P. acknowledges the Spanish Ministry of Science and Innovation for the FPI predoctoral fellowship PRE2022-103066. M.G.C. acknowledges funding from the European Union's Horizon 2020 research and innovation programme under the Marie Skłodowska-Curie grant agreement no. 101034431 (IDEAL programme).

Received: ((will be filled in by the editorial staff))
Revised: ((will be filled in by the editorial staff))
Published online: ((will be filled in by the editorial staff))